# Dirac Strings and the Nonperturbative Photon Propagator in Compact QED

KEN YEE

Department of Physics and Astronomy, LSU
Baton Rouge, Louisiana  70803-4001  USA
Email: kyee@rouge.phys.lsu.edu

ABSTRACT: In the Villain approximation of $D = 2 + 1$ compact QED, the monopole part of the partition function factorizes from the Dirac string part, which generates the photon propagator. Numerical experiments in exact compact QED confirm this result: photon mass pole $M_\gamma$, originally nonzero, is insensitive to monopole prohibition but almost vanishes if Dirac strings are prohibited.

## I. Motivation

Lattice calculations [1, 2] have evaluated the mass poles $M_{g,q,\gamma,e}$ of gluon, quark, photon, and electron propagators in LGT(lattice gauge theory). More recently, "photon" propagators in abelian-projected $SU(2)$ LGT were studied [3]. Remarkably the $SU(2)$ photon mass, as in cQED$_{3+1}$(compact QED in 3+1 dimensions) [2], is nonzero in the confine phase and vanishes in the deconfine phase. Since computing these masses, which are gauge variant [4], requires gaugefixing to either Landau or maximal abelian gauges it is unclear whether they are indicators of real physics, lattice regularization artifacts, gaugefixing artifacts [5], or some combination thereof.

The abelian-projection potentially provides a concrete framework for investigating the relationship between LGT gluon masses and confinement via abelian-projected photon masses [6]. This possibility beckons us to clarify the relationship between nonperturbative photon masses and confinement in cQED. To this end, this Letter focuses on cQED$_{2+1}$, a QCD-like model with local gauge invariance, chiral symmetry breaking, and area-law electron confinement. cQED confines because quantum monopole fluctuations [7] restrict electric flux to Abrikosov tubes of width $\lambda$ in a dual Meissner effect [8]. Recent LGT simulations resolve a finite London penetration depth $\lambda$ which, following Landau-Ginzburg theory, can be identified as the inverse mass of an effective gauge potential $A_\mu^{\text{eff}}$.

As described below, the nonperturbative zero-momentum photon propagator mass pole $M_\gamma$ in cQED$_{2+1}$ is also resolvable and nonzero. The observation of two mass scales, $\lambda$ and $M_\gamma$, raises the following question: How, if at all, is $A_\mu^{\text{eff}}$ related to primary photon field $A_\mu$ and $\lambda$ to $M_\gamma$? Are they equivalent?

Indeed, as shown below, $\lambda$ and $M_\gamma$ in cQED$_{2+1}$ are not blood relatives. The gauge-invariant quantity $\lambda$ arises from quantum monopole fluctuations while $M_\gamma$ arises from gauge-variant Dirac strings.

## II. Monopoles, Gaugefixing, and Dirac Strings

Let $\partial_\mu h_x \equiv h_{x+\hat{\mu}} - h_x$, $\partial_\mu^* h_x \equiv h_x - h_{x-\hat{\mu}}$, and $\square \equiv \partial_\mu^* \partial_\mu$. Link and plaquette angles in cQED,

$$\theta_\mu \equiv A_\mu + 2\pi n_\mu, \quad -\pi \leq A_\mu < \pi, \quad n_\mu \in \mathbf{Z} \qquad (1)$$

$$F_{\mu\nu}[A] \equiv \partial_\mu A_\nu - \partial_\nu A_\mu, \qquad (2)$$

depend only on photon $A_\mu$. The cQED action $S_c \equiv \beta \sum_{\mu<\nu}(1 - \cos F_{\mu\nu})$ is invariant under local gauge transformations

$$\delta\theta_\mu \equiv -\partial_\mu \omega_x, \quad \delta A_\mu \equiv (A_\mu - \partial_\mu \omega_x)\mathrm{Mod}[-\pi, \pi) - A_\mu. \qquad (3)$$

While plaquette $\exp(iF_{\mu\nu})$ is gauge-invariant, a gauge transformation inducing unequal shifts of $n_\mu$ on the four links of $F_{\mu\nu}$ shifts $F_{\mu\nu}$ by a $2\pi$ multiple. $F_{\mu\nu}$ decomposes into a gauge-invariant part $\Theta_{\mu\nu} \in [-\pi, \pi)$ and a gauge-variant integral kink $N_{\mu\nu} \in \mathbf{Z}$ such that

$$F_{\mu\nu} \equiv (\Theta + 2\pi N)_{\mu\nu}, \qquad (4)$$

$$\delta F_{\mu\nu} = 2\pi \delta N_{\mu\nu} = \partial_\mu(\delta\theta - \delta A)_\nu - (\mu \leftrightarrow \nu). \qquad (5)$$

Because $\cos F_{\mu\nu} = \cos \Theta_{\mu\nu}$ in $S_c$, required since $N_{\mu\nu}$ is gauge-variant, $\frac{\beta}{4}F^2$ would be a misleading approximation to $S_c$. More valid is Villain's periodic gaussian approximation $S_c \to S_c^V$ where

$$\mathrm{e}^{-S_c^V} \equiv \sum_{\{N\}} \mathrm{e}^{-\frac{\beta}{4}\sum_x (F[A] - 2\pi N)^2}, \qquad (6)$$

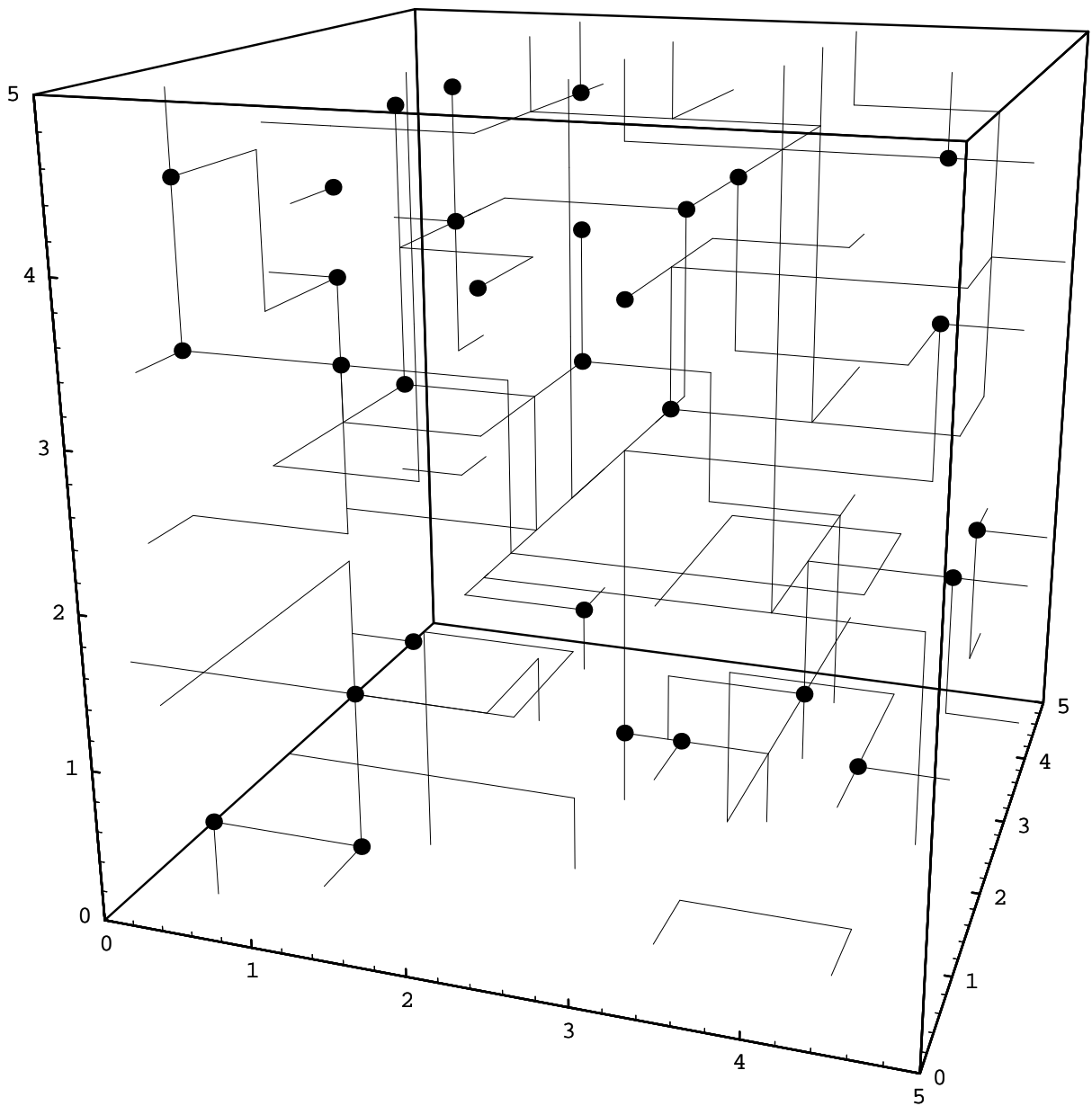

Figure 1A. Monopoles(•) and Dirac strings(—) on a typical periodic $5^3$ lattice before gaugefixing.

$\sum_{\{N\}} \equiv \prod_{\mu,\nu} \sum_{N_{\mu\nu}=-\infty}^{\infty} \delta(-N_{\nu\mu}, N_{\mu\nu})$, and $F^2 \equiv \sum_{\mu,\nu} F_{\mu\nu}^2$. The sum over $\{N\}$ in (6) is necessary to maintain gauge invariance under (5). Within an $N_{\mu\nu}$ potential well the Maxwell equation is $\partial_\mu^* F_{\mu\nu} = K_\nu \equiv 2\pi \partial_\mu^* N_{\mu\nu}$. Conserved current $K_\mu$ acts as a virtual source of electromagnetic charge. Quantum $N_{\mu\nu}$ fluctuations can lead, via the Maxwell equation, to a nonzero effective $M_\gamma$ if

$$\langle \mathcal{O}_\mu K_\mu \rangle_N = M_\gamma^2 \langle \mathcal{O}_\mu A_\mu \rangle_N + \cdots \qquad (7)$$

In (7) $\langle \ \rangle_N$ refers to the Boltzmann average over $N_{\mu\nu}$ and operator $\mathcal{O}_\mu$ serves to absorb $D = 2+1$ rotational invariance.

While $N_{\mu\nu}$ is gauge variant, spatial combinations of them form gauge-invariant structures which influence $\Theta_{\mu\nu}$ as follows. According to the Hodge-DeRham theorem

$$\Theta_{\mu\nu} = \epsilon_{\mu\nu\alpha} \partial_\alpha^* \phi + \partial_\mu \alpha_\nu - \partial_\nu \alpha_\mu, \qquad (8)$$

$$N_{\mu\nu} = \epsilon_{\mu\nu\alpha} \partial_\alpha^* m + \partial_\mu l_\nu - \partial_\nu l_\mu \qquad (9)$$

where $\phi, \alpha_\mu \in (-\infty, \infty)$ and $m, l_\mu \in \mathbf{Z}$. $\phi$ and $m$ are invariant under (3), $\alpha_\mu$ transforms like $\theta_\mu$, and $l_\mu$ like $(A-\alpha)_\mu/2\pi$. $\Theta_{\mu\nu}$ (and similarly $N_{\mu\nu}$) has 3 independent polarizations while $\phi$ and $\alpha_\mu$ are 4 functions because $\Theta_{\mu\nu}$ is invariant under $\delta \alpha_\mu = -\partial_\mu \omega_x$. In vector notation Eq. (4) becomes $\vec{B} = \vec{H} + 2\pi \vec{\eta}$ where the total $\vec{B}$, physical $\vec{H}$, and Dirac string $\vec{\eta}$ magnetic(actually electromagnetic) fields are

$$\vec{B} \equiv \nabla \times \vec{A}, \quad \vec{H} \equiv \nabla \phi + \nabla \times \vec{\alpha}, \quad \vec{\eta} = \nabla m + \nabla \times \vec{l}. \qquad (10)$$

It will be advantageous to recast $\vec{\eta}$ in terms of its divergence and curl

$$q \equiv \nabla \cdot \vec{\eta} = \Box m, \quad \vec{\rho} \equiv \nabla \times \vec{\eta} = \nabla(\nabla \cdot \vec{l}) - \Box \vec{l}. \qquad (11)$$

Since $\nabla \cdot \vec{B} = 0$ by (10), $\nabla \cdot \vec{H} = -2\pi q$, that is, $q$ causes dislocations in the physical field $\vec{H}$. By tautology $q$ is the magnetic monopole density, gauge invariant since $m$ is gauge invariant. $\vec{\rho}$, a continuous current wrapping around $\vec{\eta}$, is gauge-variant.

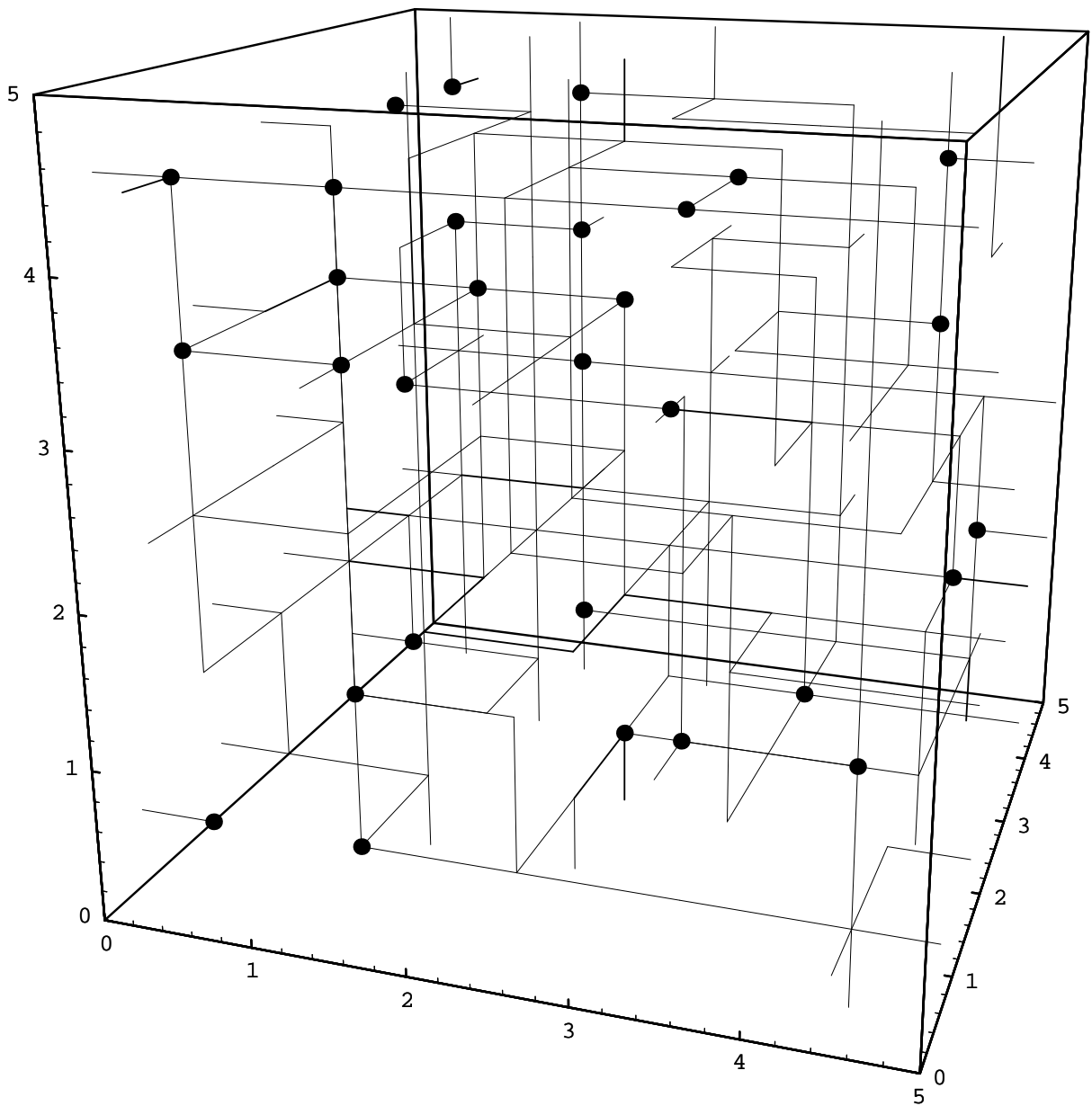

Figure 1*B*. The same lattice after Landau gaugefixing. While monopole positions do not change, there are more Dirac strings.

In general, kinks occur either in monopoles, Dirac strings connecting a monopole antimonopole pair, or Dirac string loops. Loops can either be homologically trivial or toroidally wind around the periodic boundaries. Monopole charge density $q$ is gauge-invariant but the number of string loops and the length and shape of all strings vary with gauge. Figure 1$A$ shows a monopole and Dirac string arrangement in a representative $\beta = 1.0$, periodic $5^3$ non-gaugefixed lattice. Figure 1$B$ shows the same lattice after 5000 Landau gauge-fixing sweeps. There are typically more Dirac strings after Landau gaugefixing than before, as depicted.

### III. Factorization of Photon Propagator from Monopoles

Upon adopting the condition $\nabla \cdot \vec{l} = 0$ and ignoring Laplacian zero modes, Eqs. (4)-(11) constitute 1-to-1 variables changes $\{N\} \to \{m, \vec{l}\} \to \{q, \vec{\rho}\} \to \{\phi, \vec{\alpha}\}$ where, if $\Box \Delta_{x,y} = -\delta_{x,y}$, $\partial_\mu^* \Delta^{\mu\nu} = 0$, and $\Box \Delta_{x,y}^{\mu\nu} = -\delta_{\mu,\nu}\delta_{x,y}$, then $\phi = 2\pi \sum_y \Delta_{x,y} q_y$ and $\alpha_\mu = A_\mu - 2\pi \sum_y \Delta_{x,y}^{\mu\nu} \rho_{\nu y}$. Following (4), (6), and (8)

$$Z_c^V \equiv \int_A e^{-S_c^V} = Z_m[0] \times Z_{Al}[0], \qquad (12)$$

$$Z_m[\xi] \equiv \sum_{\{q\}} e^{\sum_x \xi_x q_x - 2\pi^2 \beta \sum_{x,y} q_x \Delta(x-y) q_y}, \qquad (13)$$

$$Z_{Al}[0] = \int_\alpha e^{-\frac{\beta}{4} \sum_x F^2[\alpha]}, \quad \alpha_\mu \in (-\infty, \infty). \qquad (14)$$

Eq. (12) relies on the quadratic character of the Villain approximation and $\sum_x \epsilon_{\mu\nu\lambda} F_{\mu\nu} \partial_\lambda^* \phi = 0$; $\mathcal{O}(\Theta^4)$ corrections from the cos $\Theta$s of $S_c$ would ruin factorization. (14) says $\vec{\alpha}$ is a massless noncompact vector particle.

The dilute gas expansion and occupation number resummation over $q \in \{0, \pm 1\}$ of $Z_m$ in (13) yields [7]

$$Z_m[\xi] \propto \int_\Phi e^{-\frac{1}{4\pi^2\beta}\sum_x (\nabla(\Phi-\xi))^2 - 2\lambda^{-2}\cos\Phi} \tag{15}$$

where $\lambda^2 = 2\pi^2\beta e^{-2\pi^2\beta\Delta(0)}$. Scalar $\Phi$ is semiclassically identified with $\phi$ in (8) because, since $\sum_x \xi_x q_x = \frac{1}{2\pi}\sum_x \partial_\mu \xi_x \partial_\mu^* \phi_x$, comparing (13) to (15) yields

$$\frac{1}{2\pi}\langle\partial_\mu\phi\rangle_m = \frac{1}{Z_m}\frac{\delta Z_m}{\delta\partial_\mu^*\xi}\bigg|_{\xi\to 0} = \frac{1}{2\pi^2\beta}\langle\partial_\mu\Phi\rangle_\Phi. \tag{16}$$

Following (10) this identification yields

$$\langle\nabla\cdot\vec{H}\rangle_{Alm} = -2\pi\langle q\rangle_m = 0, \tag{17}$$

$$\langle\vec{H}_y\vec{H}_x\rangle_{Alm} = \langle(\nabla\times\vec{\alpha})_y(\nabla\times\vec{\alpha})_x\rangle_{Al} + \frac{4\pi^2}{Z_m}\frac{\delta^2 Z_m}{\delta\nabla\xi_y\delta\nabla\xi_x}. \tag{18}$$

If $\beta/\lambda \ll 1$, (17) and (18) are reproduced by an $M_\gamma = \lambda^{-1}$ *free* photon $\vec{A}^{\text{eff}}$ with $\vec{H}^{\text{eff}} \equiv \nabla\times\vec{A}^{\text{eff}}$, that is, $\nabla\cdot\vec{H}^{\text{eff}} = 0$ and $\langle\vec{H}_y^{\text{eff}}\vec{H}_x^{\text{eff}}\rangle_{\text{eff}} \approx \langle\vec{H}_y\vec{H}_x\rangle_{Alm}$. The latter relation relies on the masslessness of $\vec{\alpha}$ in (18), a consequence of (14), and $\cos\Phi \to -\Phi^2/2$ in (15). $\vec{A}^{\text{eff}}$ is the massive Landau-Ginzburg photon and $\lambda$ the London penetration depth.

The $\vec{A}$ propagator is generated by $Z_{Al}[J]$, defined by adding $J\cdot A$ to the action in (6), which does not affect factorization (12). Thus the $\vec{A}$ mass $M_\gamma$ has nothing to do with monopoles $q$ and, hence, $M_\gamma$ is not directly related to $\lambda$. A remaining logical possibility, which we do not exclude, is that $\lambda$ and $M_\gamma$ are coincidentally related via the statistical mechanics of monopoles and $\vec{\rho}$ loop gasses. $\vec{A}$, unlike $\vec{\alpha}$, may be massive because $J\cdot A$ breaks the pure $\vec{\alpha}$-dependent form of $Z_{Al}[0]$. Manipulations like those leading to (13) yield

$$Z_{Al}[J] = \int_A \sum_{\{\vec{\rho}\}} e^{\sum_x (J+\pi\beta\rho)\cdot A - \frac{\beta}{4}F^2 - \pi^2\beta\sum_y \rho\cdot\Delta\cdot\rho}.$$

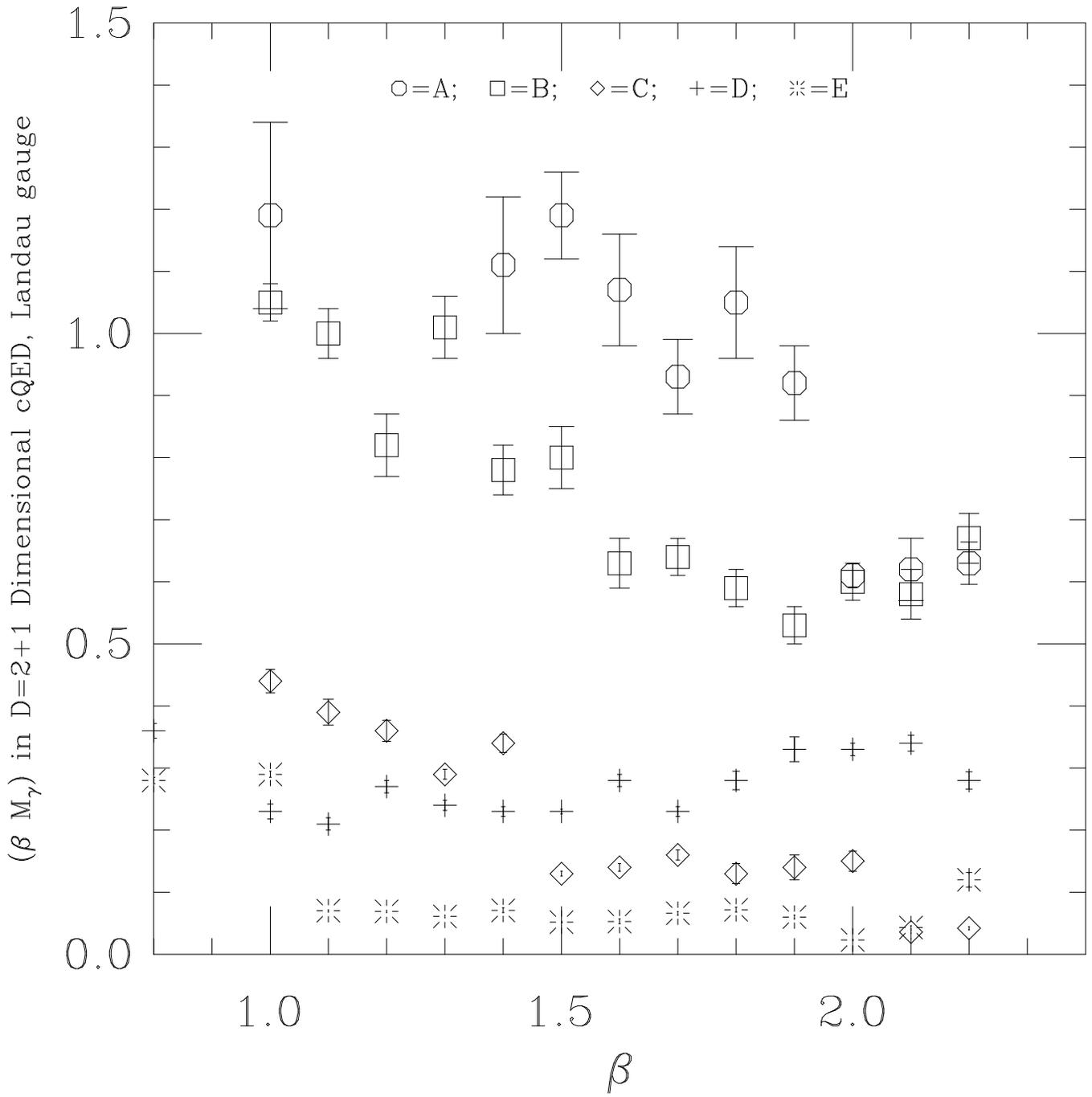

Figure 2. (A)cQED with action $S_c$;

(B)cQED with monopoles forbidden;

(C)cQED with kinks forbidden;

(D)quadratic approximation $S_c \to S_E$, kink-creating gauge transformations allowed;

(E)quadratic approximation $S_c \to S_E$, kink-creating gauge transformations forbidden.

$Z_{Al}$ is the partition function of a Coulombic $\vec{\rho}$ loop gas, "loop" since $\nabla \cdot \vec{\rho} = 0$. Interestingly $\vec{\rho}$ is a mixed state in the gas since for $M_\gamma$ to be nonzero $\sum_{y,y'} \Delta_{x,y}^{\mu\alpha} \Delta_{0,y'}^{\nu\beta} \langle \rho_{y,\alpha} \, \rho_{y',\beta} \rangle_{\vec{\rho}}$ must have a negative norm massless mode to cancel the $\alpha$ pole and an independent $M_\gamma$ mode. Note that even though photons $A_\mu$ are uncharged they suffer confinement since, heuristically, the "adjoint" Wilson loop obeys $\langle \prod_{l \in \text{loop}} A_l \rangle \approx \langle \prod_{l \in \text{loop}} \sin \theta_l \rangle \propto \text{Re} \langle \prod_{l \in \text{loop}} e^{i\theta_l} \rangle$, where cross terms are suppressed by gauge invariance.

## IV. Numerical Experiments

We have shown that the $A_\mu$ propagator decouples from monopoles in the Villain approximation and, accordingly, $M_\gamma$ is not a consequence of monopoles. Numerical experiments summarized in Figure 2 support this result in exact cQED$_{2+1}$. $\beta \times M_\gamma$, a dimensionless number in $D = 2 + 1$, is the log of the ratio of successive $\vec{p} = 0$ photon propagator timeslices. The numerical photon operator and gauge condition are $S_\mu \equiv \sin \theta_\mu$ and $\partial_\alpha^* S_\mu = 0$. The computation is based on 500 configurations on $17^2 \times 19$ lattices. The first configuration—independently for each $\beta$ value—is thermalized by 500 forty-hit, 40%-acceptance Metropolis sweeps and 5000 checkerboard gaugefixing sweeps. Configurations thereafter are separated by 5 forty-hit Metropolis sweeps and 5000 checkerboard gaugefixing sweeps. Errors are jackknife sigmas based on 10 450-configuration subaverages. Configurations $1 - 50$ are omitted from the first subaverage, $51 - 100$ from the second, and so forth.

"A" in Figure 2 refers to cQED in Landau gauge; "B" to Landau gauge cQED with monopoles prohibited [9]; "C" and "E" to cQED with kinks prohibited. Monopoles prohibition is implemented by starting with the $\theta_\mu = 0$ configuration and linkwise forbidding monopole-creating updates with the insertion of $\prod_{\{x\}} \delta_{q,0}$ into the link measure. Kinks are prohibited either by inserting $\prod_{\{N\}} \delta_{N,0}$ into the link measure("C") or by replacing $\cos F_{\mu\nu}$ in $S_c$ with $S_E \equiv \frac{\beta}{4} F_{\mu\nu}^2$ ("E") where (2) defines $F_{\mu\nu}$.

Landau gaugefixing, which cannot change $q$, proceeds normally in B. Since maintaining kink prohibition disallows kink-creating gauge transformations which may be necessary to achieve Landau gauge, a good Landau gauge is not achieved in C and E. Nontheless the photon propagator is easily resolved in all cases and, in agreement with factorization (12), $M_\gamma$ is relatively insensitive to monopole prohibition but very sensitive to kink prohibition. The small residual $M_\gamma$ in C is attributable to the $\mathcal{O}(\Theta_{\mu\nu}^4)$ terms in $S_c$ which ruin factorization (12). "D" is also based on the action $S_E$. Unlike in E, kink-creating gauge transformations are permitted during Landau gaugfixing in D. From the $S_E$ standpoint D, corrupted by action-changing kink creation and annihilation, is gauge *in*equivalent to E. The difference between $M_\gamma$ in D and E, gauge equivalent from the $S_c$ viewpoint, is an indication of how much kinks generated by the Landau gaugefixing algorithm contribute to $M_\gamma$. The smallness of $M_\gamma$ in D suggests that the kinks responsible for $M_\gamma$ in A are present in the pre-gaugefixing configurations and not created during Landau gaugefixing.

$M_\gamma$ and kink-number density $\rho$, two gauge-variant quantities, are correlated. In cases A-E at $\beta = 1.8$ in Landau gauge, $\rho_A = .41(.01)$, $\rho_C \equiv 0$, $\rho_D = .23(.004)$, and $\rho_E \sim 10^{-5}$. Since the $\beta = 1.8$ monopole number density is $8.0(1.1)10^{-3}$, forbidding monopoles doesn't change the kink density and $\rho_B = \rho_A$. In the same $\beta$ range the Axial gauge $\rho_A$ is smaller than the Landau gauge $\rho_A$ by $\sim 50\%$ and $M_\gamma$ smaller by $\sim 70\%$.

Acknowledgements

I have benefitted from discussions on aspects of monopoles, superconductivity, and effective actions with Dana Browne, Vandana Singh, H.R. Fiebig, Lai-Him Chan, Claude Bernard, and especially Dick Haymaker, who inspired me to think about the London relation. KY is supported by DOE grant DE-FG05-91ER40617; computation was at the LSU Concurrent Computing Lab.

REFERENCES

1. J. Mandula and M. Ogilvie, Phys. Lett. B185 (1987) 127; C. Bernard, D. Murphy, A. Soni, K. Yee, Nucl. Phys. B(Proc. Suppl.)17 (1990) 593; A. Nakamura and R. Sinclair, Phys. Lett. B243 (1990) 396.

2. P. Coddington, A. Hey, J. Mandula, M. Ogilvie, Phys. Lett. B197 (1987) 191.

3. P. Cea and L. Cosmai, Bari preprint BARI-TH89/91, unpublished.

4. M. Ogilvie, Phys. Lett. B231 (1989) 161; C. Bernard, A. Soni, K. Yee, Nucl. Phys. B(Proc. Suppl.)20 (1991) 410; K. Yee, BNL preprint #47712, submitted to Nucl. Phys. B[FS].

5. Ph. de Forcrand, J. Hetrick, A. Nakamura, M. Plewnia, Nucl. Phys. B(Proc. Suppl.)20 (1991) 194; S. Petrarca, C. Parrinello, A. Vladikas, M. Paciello, B. Taglienti, Nucl. Phys. B26 (1992) 435.

6. K. Yee, work in progress.

7. A. Polyakov, Phys. Lett. 59B (1975) 82; T. Banks, R. Myerson, and J. Kogut, Nucl. Phys. B129 (1977) 493; G. 't Hooft, Nucl. Phys. B190 (1981) 455; J. Smit and A. van der Sijs, Nucl. Phys. B355 (1991) 603.

8. M. Peskin, Ann. Phys. 113 (1978) 122; V. Singh, R. Haymaker, D. Browne, to be published in Phys. Rev. D.

9. T. DeGrand and D. Toussaint, Phys. Rev. D22 (1980) 2478; J. Barber and R. Shrock, Phys. Lett. 152B (1985) 221.